\documentclass[a4paper,twocolumn,superscriptaddress, aps, floatfix]{revtex4-1}

\usepackage{epsfig}
\usepackage{verbatim}	 
\usepackage{amssymb}
\usepackage{bm}
\usepackage{amsmath}
\usepackage{psfrag}
\usepackage{color}
\usepackage[latin1]{inputenc}
\usepackage[english]{babel}
\usepackage{graphicx}
\usepackage{braket}
\usepackage{eucal}
\usepackage{natbib}
\usepackage{hyperref}
\usepackage{lipsum}
\usepackage{mathtools}
\usepackage{multirow}
\usepackage{cleveref}
\usepackage[colorinlistoftodos]{todonotes}

\addto\captionsenglish{}

\newcommand{\MT}[1]{\textcolor{red}{(Melvyn: #1)}}

\begin{document}

\author{Fabian Baumann}
\thanks{Corresponding author: fabian.olit@gmail.com}
\affiliation{Institut f\"ur Physik, Humboldt-Universit\"at zu Berlin, Newtonstra\ss e 15, 12489 Berlin, Germany}

\author{Igor M. Sokolov}
\affiliation{Institut f\"ur Physik, Humboldt-Universit\"at zu Berlin, Newtonstra\ss e 15, 12489 Berlin, Germany}
\affiliation{IRIS Adlershof, Humboldt-Universit\"at zu Berlin, Zum Gro\ss en Windkanal 6, 12489, Berlin, Germany}

\author{Melvyn Tyloo}
\address{School of Engineering, University of Applied Sciences of Western Switzerland HES-SO, CH-1951 Sion, Switzerland.}

\title{
Periodic Coupling inhibits Second-order Consensus on Networks}
\begin{abstract}
Consensus algorithms on networks have received increasing attention in recent years for various applications ranging from animal flocking to multi-vehicle co-ordination. Building on the established model for second-order consensus of multi-agent networks, we uncover a mechanism inhibiting the formation of collective consensus states via rather small time-periodic coupling modulations. We treat the model in its spectral decomposition and find analytically that for certain intermediate coupling frequencies parametric resonance is induced on a network level -- at odds with the expected emergence of consensus for very short and long coupling time-scales. Our formalism precisely predicts those resonance frequencies and links them to the Laplacian spectrum of the static backbone network. The excitation of the system is furthermore quantified within the theory of parametric resonance, which we extend to the domain of networks with time-periodic couplings.
\end{abstract}

\maketitle
\section{Introduction}
\begin{figure*}\includegraphics[width=\linewidth]{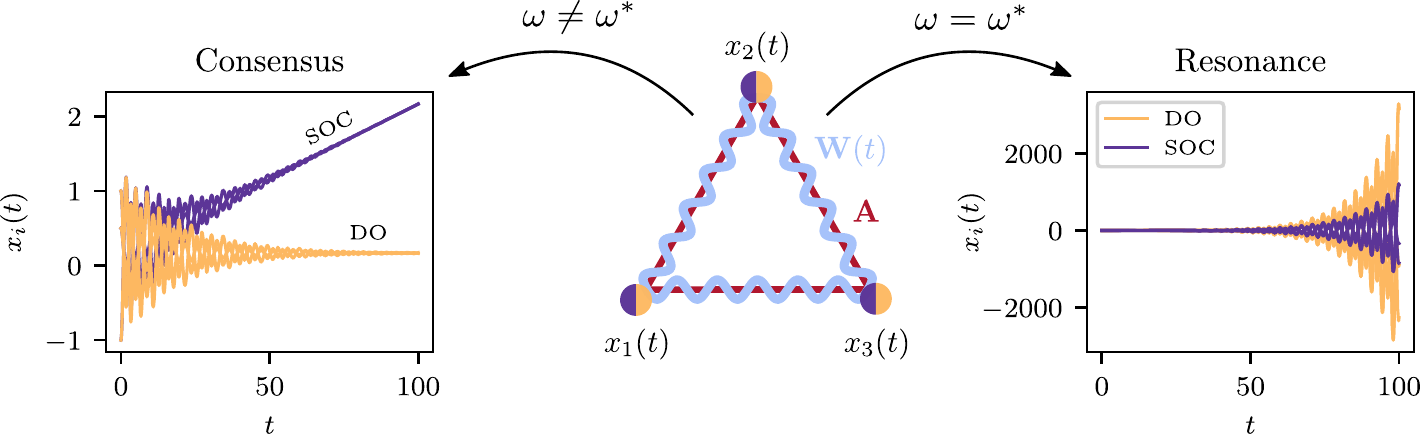} \caption{\textbf{Dynamics of the DO and SOC model} for resonant and off-resonant coupling modulation. For ease of illustrations we consider a small system of three agents connected via a complete graph $\mathbf{A}$. The coupling strength between any two agents is modulated time-periodically according to $\mathbf{W}(t)=f(t)\mathbf{A}$, cf. Eq.~\eqref{eq:modulation}. In the case of a static network $\mathbf{A}$ for $\omega=0$, or sufficiently off-resonant modulations ($\omega\neq\omega^*$), the agents will reach a consensus in the limit of $t\rightarrow\infty$. In the case of the DO model velocities vanish and all agents come to rest in the same position. Instead, for the SOC model consensus is defined as a state in which all agents move together (being in the same place) at equal velocities.
In contrast, for resonant coupling modulations ($\omega=\omega_\alpha^*$) parametric excitation, on a collective network level, inhibits the establishment of global consensus states and the agents' amplitudes diverge, as depicted in the right panel.
}
\label{fig:fig1_example}
\end{figure*}

De-centralized collective behaviors arise in a variety of different contexts, where single units interact \emph{without} a centralized authority controlling the overall process. Often 
such interactions correspond to exchanges of information that are modeled using networks. 
Depending on the context, the information flow between two network nodes may be interpreted differently -- ranging from single neurons coupled via action potentials \cite{bassett2017network,izhikevich2007dynamical}, animal groups exchanging cues on prey or predators \cite{rosenthal2015revealing,katz2011inferring}, to communicating robots within vehicular platoons \cite{tanner2003stable,gupta2003stability,rodriguez2003cooperative,werfel2014designing}. Notably, it has been shown that for many coupled dynamical systems, the arising collective states strongly depend on the characteristics of the network topology \cite{porter2016dynamical}. 

Traditionally, special attention has been paid to the emergence of \emph{consensus} \cite{degroot1974reaching,baronchelli2018emergence,castellano2009statistical}.
The outstanding role of consensus states for many systems is often justified by collective decision making, which requires agreement on a population level to reach a collective goal \cite{baronchelli2018emergence}. For example, fish schools have been shown to reach increased levels of collective sensitivity if individuals are more strongly aligned constituting a consensus on the collective heading direction \cite{couzin2007collective}. In social systems language-based communication relies on sets of quasi-static vocabularies, where a consensus on the meaning of a word has been achieved \cite{hechter2001social}.  And in vehicular platoons, consensus is often attributed to situations, in which all involved robotic units need to be located at the same position, point in the same direction or move at the same speed \cite{tanner2003stable,gupta2003stability,rodriguez2003cooperative,ren2007information,zhang2015model}. 

Taking into account the specific physical constraints of single units (or agents) gives rise to different types of consensus algorithms. Considerable efforts have been devoted to second-order consensus protocols -- referring to the Newtonian dynamics of single agents -- both in discrete and continuous time \cite{ren2007information,ren2005second,ren2007distributed,yang2008stability,qin2012stationary,ren2007second,yu2009second,yu2010some,goldin2014consensus,ren2007distributed,ai2016second,hu2011distributed}.
Illustrative examples include applications to leader-following control \cite{hu2011distributed}, flocking \cite{zhang2015model} and mulit-vehicle co-orperative control \cite{ren2007distributed}.
More specifically, second-order consensus algorithms were investigated for limited interaction ranges \cite{ai2016second}, time delays \cite{yang2008stability} and for the case of different networks for position and velocity coupling \cite{goldin2014consensus}. Most importantly it was shown, that networks with a (directed) spanning tree do not yield a sufficient condition for achieving consensus in second-order systems \cite{ren2005consensus} -- at odds with first-order counterparts.

While the vast majority of research on consensus focused on static networks \cite{ren2007information,ren2005second,ren2007distributed,yang2008stability,yu2009second,yu2010some,goldin2014consensus,ren2007distributed,ai2016second,hu2011distributed}, 
less attention has been paid to the effects arising on temporal, i.e. time-varying networks. Exceptions include investigations of multi-agent systems on switching topologies \cite{ren2007second,qin2012stationary}, where the interaction network changes in discrete steps. Similar to the case of static networks -- and again in contrast to first-order systems -- it was found that consensus is not necessarily achieved if infinitely many unions of consecutive network snapshots have a directed spanning tree \cite{ren2007second}. Related to consensus formation is the concept of synchronization on networks of nonlinear oscillators \cite{pikovsky2003synchronization,arenas2008synchronization}, which was investigated for a further type of temporal network. Here -- in contrast to switching networks -- the coupling strengths are time-dependent, while the topology of the network does not change over time. In Refs.~\cite{li2018periodic,PhysRevE.98.012304} such network coupling was assumed to be periodically modulated and it was found, that such modulation may strongly impact the synchronization behavior of the considered systems. In Ref. \cite{PhysRevE.98.012304}, using the classical R\"ossler oscillator, network synchronizability could be significantly improved by tuning the coupling frequency. In contrast, the same periodic coupling was found to suppress synchronization in ensembles of globally coupled first-order Kuramoto oscillators \cite{li2018periodic}.


In this paper we uncover yet another mechanism inhibiting the formation of collective second-order consensus states on networks. It is induced, not by a switching network topology \cite{ren2007second,qin2012stationary}, but the time-periodic modulation of coupling strengths as proposed in  \cite{li2018periodic,PhysRevE.98.012304}. 
To this end we investigate a general model for coupled oscillators with Newtonian dynamics \cite{oral2017performance} in two different settings: i) linearly coupled damped oscillators (DO) \cite{zhan2013matching}, and ii) the classical second-order consensus model (SOC) with velocity alignment \cite{ai2016second,ren2005second,goldin2014consensus,yu2009second,yu2010some,yang2008stability,ren2007distributed}.
For both models, the interaction strengths
are periodically modulated around well defined mean values, which are encoded in a static and symmetric backbone network. In the limits of very fast and very slow modulations, we find that global consensus states emerge --  as expected for the time-averaged static connectivity \cite{ren2007distributed}. Interestingly, however, specific intermediate modulation frequencies, lead to an excitation of the system, and thereby inhibit the formation of a global consensus. 

{As we demonstrate, this collective dynamical feature can be understood in terms of parametric resonance \cite{landau1984lehrbuch} on the level of network modes,  revealed by the spectral decomposition of the system.} 
In our analysis we link the corresponding resonance frequencies (of such parametric oscillators) to the Laplacian spectrum of the static backbone network.  We furthermore estimate the growth rates of oscillation amplitudes in the case of resonance. As we will briefly discuss in the Appendix using the Kuramoto model~\cite{kuramoto1975self} with inertia~\cite{tanaka1997self,dorfler2013synchronization,Rod16}, our framework may also be applied to non-linearly coupled systems, close to a synchronization fixed point. 

{In the left panel of Fig.~\ref{fig:fig1_example} the dynamics of both considered models towards consensus is illustrated. While for the damped oscillator model (DO) a motionless consensus via dissipation (orange lines) emerges, the consensus state of the SOC model preserves the total kinetic energy, such that all agents travel as one cluster with a specific finite velocity (purple lines). In the case of resonant modulation, consensus is inhibited for both models, and the agents' amplitudes diverge, cf. right panel of Fig.~\ref{fig:fig1_example}.}

The rest of the paper is structured as follows. In Sec.~\ref{sec1} we introduce the considered general model of coupled agents. Using a spectral decomposition we derive decoupled mode equations, which can be treated within the theory of parametric oscillators. This allows to link the resonance frequencies of both, the DO and the SOC model, to the Laplacian spectrum of the backbone matrix. In Sec.~\ref{sec:results} we validate and complement our theoretical considerations using numerical simulations on different networks, and briefly discuss the theory's applicability to non-linearly coupled oscillators, cf. App.~\ref{sec:non-lin-osc}. The work is summarized and concluded in Sec.~\ref{sec:conclusions}. 

\section{Model and theory}\label{sec1}
Let us consider a network of $n$ coupled units (or agents). The coupling matrix, $\mathbf{W}(t)=f(t)\mathbf{A}$, which encodes the time-dependent connection strengths between two agents is composed of two parts: i) a static and symmetric backbone network, described by the adjacency matrix $\mathbf{A}$, and ii) the periodic modulation  \cite{li2018periodic,PhysRevE.98.012304}
\begin{equation}\label{eq:modulation}
    f(t) = 1+h\sin(\omega t)\,,
\end{equation}
with frequency $\omega$ and amplitude $h$. If two agents $i$ and $j$ are connected the matrix elements $A_{ij}(=A_{ji})=1$ and $A_{ij}(=A_{ji})=0$, otherwise.  
{To exclude repulsive coupling we assume $h\in[0,1]$ in the following.} 
Note, that due to the periodicity of $f(t)$, the network backbone and the coupling matrix are connected via the temporal average $\mathbf{A}=\lim_{t\rightarrow\infty}T^{-1}\int_0^T\mathbf{W}(t) {\rm d}t$.
Each agent $i$ is associated with a distinct node of the network and is characterized by a time-dependent state variable, $x_i(t)$, where we will omit the time dependence for brevity in the following. 

To model the collective dynamics we consider the established model for second-order consensus on networks \cite{ai2016second,ren2005second,goldin2014consensus,yu2009second,yu2010some,yang2008stability,ren2007distributed}, where we additionally assume time-periodic couplings. {In its most general form  including dissipation \cite{Gru18}, it is described by the following set of second-order differential equations}
\begin{align}\label{eq:dyn}
\ddot{x}_i + d(t)\,\dot{x}_i = &  -\gamma\sum_{j=1}^{n} W_{ij}(t)\,(x_i-x_j)\nonumber\\&-\mu\sum_{j=1}^{n} W_{ij}(t)\,(\dot{x}_i-\dot{x}_j)\,,
\end{align}
where $W_{ij}(t)$ is the $ij$--th element of the symmetric coupling matrix $\mathbf{W}$ at time $t$
and $d(t)$ denotes the damping coefficient.
The coefficients $\gamma,\mu>0$ control the strengths of diffusive coupling between the agents' positions [first sum on the r.h.s. of Eq.~\eqref{eq:dyn}] and velocities (second sum), respectively. 

In the first case, the DO model, we assume that each agent dissipates energy $(d>0)$, and agents are not coupled with respect to their velocities ($\mu=0$). This corresponds to an ensemble of $n$ one-dimensional harmonic oscillators, coupled via linear springs with a time-dependent spring constant $\gamma W_{ij}(t)$. Similar harmonic oscillators with time-invariant couplings have been investigated on networks using the classical mass spring model \cite{zhan2013matching}. More importantly, however, the DO model describes the linearized dynamics of non-linearly coupled oscillators, such as the second-order Kuramoto model, which was previously used to investigate synchronization phenomena on power-grids \cite{dorfler2013synchronization,Rod16}. In Appendix \ref{sec:non-lin-osc}, we will demonstrate that the developed formalism may also be applied to such non-linearly coupled systems. Note, that the agents in the damped oscillator (DO) model achieve consensus due to the finite dissipation of $d>0$ and the diffusive position coupling. Hence, the system evolves such that all agents will finally come to rest at the same position $c$ and consensus is characterized as $x_i=c\;,\forall i$ and $\dot{x}_i(t\rightarrow\infty)=0,\;\forall i$. The dynamics is illustrated in the left panel of Fig.~\ref{fig:fig1_example}.

Subsequently, we investigate the effects of periodic coupling for the classical model for second-order consensus (SOC) on networks \cite{ren2007information,ren2005second,ren2007distributed,yang2008stability,qin2012stationary,ren2007second,yu2009second,yu2010some,goldin2014consensus,ren2007distributed,ai2016second,hu2011distributed}, where agents do not dissipate energy ($d=0$) but may establish consensus due to a velocity alignment mechanism, for $\mu>0$, cf. Eq.~\eqref{eq:dyn}. Here, consensus is usually defined as situations with $|x_i-x_j|=0\;, \forall i,j$ and $|\dot{x}_i-\dot{x}_j|=0\;, \forall i,j$ -- a state with equal but potentially finite velocities \cite{ren2007distributed}.  {In the following, we will show that both models are susceptible to time-periodic coupling, which inhibits the formation of consensus states. The presented formalism furthermore reveals an intimate relation between dissipation and velocity alignment and their roles in consensus formation.} 

To proceed further, we express the general model, Eqs.~\eqref{eq:dyn}, 
in vectorial form as 
\begin{align}\label{eq:vectorial}
\Ddot{\bm x} + d(t)\dot{\bm x} = -\mathbb{L}(t)(\gamma{\bm x}+\mu\dot{{\bm x}}) \;,
\end{align}
where $\mathbb{L}(t)$ denotes the time-dependent network Laplacian of the system. Analogous to the temporal coupling matrix, $\mathbf{W}(t)$, $\mathbb{L}(t)$ factorizes into the time-independent Laplacian $\mathbb{L}^{(0)}$ and the coupling modulation function $f(t)$, i.e. $\mathbb{L}(t)=\mathbb{L}^{(0)}f(t)$. The elements of $\mathbb{L}^{(0)}$ are obtained from the static backbone connectivity ${\mathbf{A}}$ as  
\begin{equation}\label{eq:laplacian}
{\mathbb L}^{(0)}_{ij} = 
\left\{ 
\begin{array}{cc}
-{A}_{ij}  \, , & i \ne j \, , \\
\sum_k {A}_{ik}  \, , & i=j \,.
\end{array}
\right.
\end{equation}
Note, that the modulation $f(t)$ acts upon every connection of the static backbone network ${\mathbf{A}}$ simultaneously \cite{li2018periodic,PhysRevE.98.012304}. Therefore, also the time-dependent eigenvalues of $\mathbb{L}(t)$, $\lambda_\alpha(t)$, can be expressed as  $\lambda_\alpha(t)=\lambda_\alpha^{(0)}f(t)$ where $\{\lambda_\alpha^{(0)}\}$ is the set of eigenvalues of $\mathbb{L}^{(0)}$. 

A spectral decomposition of Eq.~\eqref{eq:vectorial} using $\mathbf{x}(t)=\sum_\alpha c_\alpha(t)\,\mathbf{u}_\alpha$, involving the eigenvectors of $\mathbb{L}^{(0)}$, $\mathbf{u}_\alpha$, gives rise to the following second-order differential equation 
\begin{eqnarray}\label{eq:ca}
\Ddot{c}_\alpha + k(t)\dot{c}_\alpha + \gamma\lambda_\alpha(t)c_\alpha = 0 \;,
\end{eqnarray}
for the $\alpha$--th expansion coefficient, $c_\alpha$, of $\mathbf{x}(t)$ on set of eigenvectors $\{\mathbf{u}_\alpha\}$, where we defined $k(t)=d(t)+\mu\lambda_\alpha(t)$. 

In this form, the effect of a time-varying coupling on the dynamics of the system becomes particularly evident: equation~\eqref{eq:ca} describes the collective dynamics on the level of network modes and manifests itself as  
parametric oscillator \cite{turyn1993damped}. Such oscillators are
not excited by an explicit external force,
but are susceptible to
periodic variations of the system's parameters -- a phenomenon which is called parametric resonance. Therefore,  parametric oscillators cannot be driven out of a stable equilibrium, but existing deviations from such a stable point may be amplified exponentially by carefully tuned periodic modulations, via $k(t)$ or $\lambda_\alpha(t)$. 
The composition of $k(t)$ reveals the analogy between dissipation $d>0$ in the case of coupled damped harmonic oscillators and the velocity alignment for the SOC model, $\mu>0$. Both mechanisms may contribute to the damping of the $\alpha$--th mode, and will therefore influence the dynamics in similar ways.

{In what follows we treat Eq.~\eqref{eq:ca} analytically and identify, for both model versions (DO and SOC), certain resonant modulation frequencies $\omega^*$ [cf. Eq.~\eqref{eq:modulation}], for which the corresponding networked systems cannot establish a global consensus. 
At a first glance this may come as a surprise: diffusively coupled agents with $\mathbf{W}(t)>0$, generally aim to assimilate, hence, one might expect the emergence of consensus. Equation~\eqref{eq:ca}, however, suggests that periodically modulated coupling strengths translate into collective parametric resonance, which is, as we will see in the following, characterized by an exponential growth of the agents' amplitudes. Here, we establish an analytical link between the spectrum of the static Laplacian $\mathbb{L}^{(0)}$ and the expected resonance frequencies $\omega^*_\alpha$. Finally, we quantify the exponential excitation in the case of resonant modulation.} 


To obtain the set of resonance frequencies for both models we proceed by expressing Eq.~\eqref{eq:ca} in a more suitable form using the following variable transformations, 
\begin{eqnarray}\label{eq:subs1}
c_\alpha(t)&=&e^{-K(t)}q_\alpha(t)\;,\\
K(t)&=& \frac{1}{2}\int_0^t k(t'){\rm d}t'\label{eq:subs2} \;.
\end{eqnarray}
Substituting those 
into Eq.~\eqref{eq:ca} yields
\begin{eqnarray}\label{eq:qa}
\Ddot{q}_\alpha + \Omega_\alpha^2(t) q_\alpha = 0\;,
\end{eqnarray}
the Hill equation \cite{hill1886part}, with the time-dependent frequency $\Omega_\alpha(t)$, defined as  
\begin{equation}\label{eq:Omega}
\Omega_\alpha^2(t)=\gamma\lambda_\alpha(t) - k^2(t)/4 - \dot{k}(t)/2\,.
\end{equation} 
Note, that the time dependence in Eq.~\eqref{eq:Omega} either results from a periodic modulation of the damping $d(t)$, the $\alpha$--th time-dependent eigenvalue $\lambda_\alpha(t)$, or both. Up to this point, the framework includes both considered cases, which differ regarding Eq.~\eqref{eq:Omega}. Below, we will therefore discuss both models separately.

\subsection{Damped oscillator model (DO)}
In the case of the DO model we assume for simplicity a time-independent damping coefficient $d$. Then Eq.~\eqref{eq:Omega} becomes  
\begin{equation}\label{eq:Omega2}
\Omega_\alpha^2(t)=\gamma\lambda_\alpha^{(0)}[1+h\sin(\omega t)] - d^2/4,
\end{equation}
where we have used $\lambda_\alpha(t)=\lambda_\alpha^{(0)}[1+h\sin(\omega t)]$. Defining $\omega_\alpha^2=\gamma{\lambda_\alpha^{(0)}}-d^2/4$\, yields
\begin{eqnarray}\label{eq:qa2}
\Ddot{q}_\alpha + \omega_\alpha^2\left[1+\frac{\gamma\lambda_\alpha^{(0)}h}{\omega_\alpha^2}\sin(\omega t)\right] q_\alpha = 0\;,
\end{eqnarray}
such that parametric excitation for the $\alpha$--th mode is  expected for $\omega_\alpha^*$ 
\begin{equation}\label{eq:def-res}
    \omega_\alpha^*=2\omega_\alpha =2\sqrt{\gamma\lambda_\alpha^{(0)}-\left(\frac{d}{2}\right)^2}\,,
\end{equation}
corresponding to twice the natural frequency of the system $\omega_\alpha$ --  
as in the case of a one-dimensional parametrically driven oscillator \cite{landau1984lehrbuch}.
Note, that not all eigenvalues $\lambda_\alpha^{(0)}$ are expected to give rise to a resonance -- only those eigenvalues, which fulfill the following requirements:
\begin{align}\label{eq:cond}
4\gamma\lambda_\alpha^{(0)}&>&d^2 \;,
\end{align}
\begin{align}\label{eq:cond2}
\left|\frac{\gamma\lambda_\alpha^{(0)}h}{4\omega_\alpha}\right|&>&d \;.
\end{align}
Indeed, Eq.~(\ref{eq:cond}) ensures that $\omega_\alpha^*$ in Eq.~(\ref{eq:def-res}) is purely real while Eq.~(\ref{eq:cond2}) yields an exponential growth of the oscillations (see below). Interestingly, as both conditions depend on the mode index $\alpha$, only a subset of the modes may be activated by the periodic coupling depending on the parameter values of $d$ and $h$. More precisely those conditions set a lower bound under which natural frequencies cannot be activated.
Apart from locating the resonance frequencies $\omega_\alpha^*$ in the system, it is instructive to quantify the exponential growth of the $\alpha$--th mode due to resonant modulation. 

To this end, we first assume small oscillation amplitudes in Eq.~\eqref{eq:qa2}, i.e. $\gamma\lambda_\alpha^{(0)}h/\omega_\alpha^2\ll 1$\,. Then the solution of Eq.~\eqref{eq:qa2} should take the form
\begin{eqnarray}\label{eq:q(t)}
q_\alpha(t)=&a_\alpha(t)\cos[(\omega_\alpha+\varepsilon) t]\nonumber \\&+ b_\alpha(t)\sin[(\omega_\alpha+\varepsilon) t] \,.
\end{eqnarray}
where $\varepsilon$ defines the detuning from $\omega_\alpha^*$. Note, that this ansatz will not yield an exact solution to Eq.~\eqref{eq:qa2}, as terms with multiples of the frequency $(2\omega_\alpha + \epsilon)$ have been neglected, it is, however, sufficient in the case of small modulation amplitudes \cite{landau1984lehrbuch}.
Inserting the expression for $q_\alpha(t)$ into Eq.~\eqref{eq:qa2} for the particular case of $\omega_\alpha^*=2\omega_\alpha + \epsilon$\,, and neglecting terms of order $\mathcal{O}(h^2)$, yields
\begin{eqnarray}\label{eq:ab}
-2\dot{a}_\alpha + a_\alpha\frac{\gamma\lambda_\alpha^{(0)}h}{2\omega_\alpha} -2\varepsilon b_\alpha &=&0 \;,\nonumber\\
2\dot{b}_\alpha + b_\alpha\frac{\gamma\lambda_\alpha^{(0)}h}{2\omega_\alpha} -2\varepsilon a_\alpha&=&0 \;.
\end{eqnarray}
We search the solutions to Eqs.~(\ref{eq:ab}) in the form of $a_\alpha(t), \;b_\alpha(t)\sim e^{s_\alpha t}$, respectively, which yields the following expression for the exponent $s_\alpha$
\begin{align}\label{eq:s}
s_\alpha^2 &= \left[\frac{{\gamma\lambda_\alpha^{(0)}}h}{4\omega_\alpha}\right]^2-\varepsilon^2\nonumber\\
&=\left[\frac{{\gamma\lambda_\alpha^{(0)}}h}{4\sqrt{{\gamma\lambda_\alpha^{(0)}}-d^2/4}}\right]^2-\varepsilon^2\,.
\end{align}
Hence, resonance for the $\alpha$--th mode is expected to appear on the following interval of $\epsilon$,
\begin{equation}\label{eq:epsilon_range}
 - \frac{{\gamma\lambda_\alpha^{(0)}}h}{4\sqrt{{\gamma\lambda_\alpha^{(0)}}-d^2/4}}  < \varepsilon < \frac{{\gamma\lambda_\alpha^{(0)}}h}{4\sqrt{{\gamma\lambda_\alpha^{(0)}}-d^2/4}}\,, 
\end{equation}
centered around the resonance frequency $\omega_\alpha^*$. Note, that the width of the interval depends linearly on $h$, such that the excitation region is increased for larger modulation amplitudes. In the case of perfectly resonant modulation ($\varepsilon=0$) the amplitudes $a_\alpha(t)$ and $b_\alpha(t)$, respectively, are expected to grow exponentially as $e^{s_\alpha t}$\,. Re-substituting this into the initial expression for the expansion coefficients yields $c_\alpha(t)\propto e^{(s_\alpha-d)t}$, from which condition Eq.~(\ref{eq:cond2}) is derived.

\subsection{Second-order consensus model (SOC)}
In a similar manner we may approach the model of second-order consensus, which will be discussed in the following. Here, we consider Eq.~\eqref{eq:vectorial} without damping ($d=0$) and with diffusive velocity coupling ($\mu>0$), as an established approach for second-order consensus.

Starting again from Eq.~\eqref{eq:Omega} {for $k(t)=\mu\lambda_\alpha^{(0)}[1+h\sin(\omega t)]$}, we get
\begin{align}\label{eq:qa_extension3}
\Omega_\alpha^2(t) \quad= &\quad\,\, \gamma\lambda_\alpha^{(0)}-\left(\frac{\mu\lambda_\alpha^{(0)}}{2}\right)^2\\&+\left(\gamma\lambda_\alpha^{(0)}h-2h\left(\frac{\mu\lambda_\alpha^{(0)}}{2}\right)^2\right)\sin(\omega t)\nonumber\\&-\frac{h\omega\mu\lambda_\alpha^{(0)}}{2}\cos(\omega t)-\left(\frac{h\mu\lambda_\alpha^{(0)}}{2}\right)^2\sin^2(\omega t)\,.\nonumber
\end{align}
Small modulation amplitudes, $h\ll1$, allow to neglect terms of the order $\mathcal{O}(h^2)$ in the expression for $\Omega_\alpha^2(t)$. Combining the linear cosine and sine terms, then yields
\begin{align}\label{eq:qa_extension3}
\Omega_\alpha^2(t) 
\simeq \omega^2_\alpha \Bigg(1+\frac{B}{\omega_\alpha^2}\sin(\omega t+\varphi)\Bigg)
\end{align}
with 

\begin{equation}\label{eq:omega_SOC}
    \omega_\alpha^2=\gamma\lambda_\alpha^{(0)}-\left(\frac{\mu\lambda_\alpha^{(0)}}{2}\right)^2\,,
\end{equation}

\begin{equation}
    B= h\sqrt{\left(\frac{\omega\mu\lambda_\alpha^{(0)}}{2}\right)^2+\left(\gamma\lambda_\alpha^{(0)}-\frac{\left(\mu\lambda_\alpha^{(0)}\right)^2}{2}\right)^2}
\end{equation}
and 
\begin{equation}
    \varphi = \tan^{-1}\left(\frac{\omega\mu\lambda_\alpha^{(0)}}{2\gamma\lambda_\alpha^{(0)}-\left(\mu\lambda_\alpha^{(0)}\right)^2}\right)\,.
\end{equation}
As in the previous case of the DO model, the resonance frequency corresponds to twice the natural frequency of the system $\omega_\alpha$, i.e.
\begin{equation}\label{eq:res-ren-attkins}
    \omega_\alpha^*=2\omega_\alpha =2\sqrt{\gamma\lambda_\alpha^{(0)}-\left(\frac{\mu\lambda_\alpha^{(0)}}{2}\right)^2}\,
\end{equation}
and we get 
\begin{align}\label{eq:s_alpha_SOC}
s_\alpha^2 &= \left[\frac{B}{4\omega_\alpha}\right]^2-\varepsilon^2\,,
\end{align}
where $c_\alpha(t)$ is expected to grow exponentially if $s_\alpha t-K(t)>0$. The detuning range for the SOC is limited to the interval $-B/(4\omega_\alpha)<\epsilon<B/(4\omega_\alpha)$ around $\omega_\alpha^*$. Evaluating the integral in Eq.~\eqref{eq:subs2} yields  
\begin{align}
K(t) 
     =&\frac{\mu\lambda_\alpha^{(0)}}{2}\Big(t+\frac{h}{\omega}(1-\cos(\omega t))\Big)\,,
\end{align}
which reduces for $h\ll1$ to 
\begin{equation}
K(t)\simeq \frac{\mu\lambda_\alpha^{(0)}}{2}t\,.
\end{equation}
Resonances are therefore expected if the following two inequalities are fulfilled: 
\begin{align}\label{eq:condition1_SOC}
    B\omega_\alpha-2\mu\lambda_\alpha^{(0)}>0
\end{align}
and
\begin{equation}\label{eq:condition2_SOC}
    \gamma - \frac{\mu^2}{4}\lambda_\alpha^{(0)} >0\,.
\end{equation}
Note, that as in the case of the DO model, the velocity alignment, controlled by the parameter $\mu$, {is hampering the emergence of parametric resonance and therefore fosters consensus.}

In Fig.~\ref{fig:fig1_example} the two different situations, consensus and resonance are illustrated for both models. In each case the basic model setup consists of the same static backbone network, an all-to-all network of three nodes.
The pairwise coupling strengths, $W_{ij}(t)$, are either modulated off-resonantly ($\omega\ll\omega^*$) (left panel) or in a perfectly resonant manner as  $\omega=\omega^*$ (right panel). While the system approaches a global consensus in the former case, consensus is inhibited by parametric excitation for resonant modulation. Instead, the nodes' amplitudes grow exponentially and do not reach a steady state where $|x_i-x_j|=0\;\forall\;i,j$.

The presented results are rather general. They allow to determine the resonance frequencies $\omega_\alpha^*$ for systems on arbitrary static backbone structures ${\mathbf{A}}$, where the coupling between agents is modulated by a periodic function $f(t)$ {of small amplitudes}. In the following we investigate those resonance effects for both models on different backbone networks. While the eigenvalues $\lambda_\alpha^{(0)}$ of $\mathbb{L}^{(0)}$, required for the determination of $\{\omega^*_\alpha\}$, can be obtained analytically for simple regular graphs, one needs to determine them numerically for more complex or random networks. 

\begin{figure}\includegraphics[width=\linewidth]{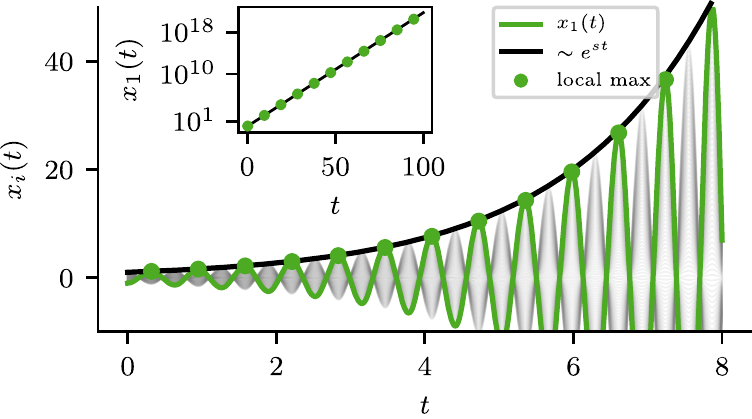} 
\caption{\textbf{Amplitude growth for resonant modulation} with $\omega=\omega^*$ and $d=\mu=0$. The exponential growth of the agents' amplitudes is quantitatively captured by the time-evolution of the $\alpha$--th expansion coefficient $c_\alpha(t)\sim e^{st}$\, (black line). The green trajectory corresponds to the time evolution of a single agent $i$, and its local maximum values are depicted as green dots. The inset shows the exponential growth as a subset of local maximum values of the same agent (green dots) and $c_\alpha(t)$ (black line) over many orders of magnitude. The shown results are for a complete graph with $n=100$ nodes.} \label{fig:single_exp_growth}
\end{figure}

\section{Numerical results}\label{sec:results}
In order to validate the theoretical analysis we perform numerical simulations. 
As suggested by the theory, the agents in both models will or will not reach a global consensus, depending on the modulation frequency $\omega$. To detect the corresponding ranges of resonance numerically we shall proceed as follows. 

We integrate Eqs.~\eqref{eq:vectorial} using a 4th--order Runge-Kutta scheme \cite{press1989numerical}, by which we obtain the time evolution of the system, $\mathbf{x}(t)$, for a specific modulation frequency $\omega$. Averaging the integral $\int \mathbf{x}^2(t)\,{\rm d}t$, obtained for each modulation frequency $\omega$, over multiple simulation runs with random initial conditions (IC), yields
\begin{equation}
    \mathcal{A}(\omega)=\left\langle\int\limits_0^{T}\mathbf{x}^2(t)\,{\rm d}t\right\rangle_{\{\mathrm{IC}\}}\,,
\end{equation}
which quantifies the excitation of the system in terms of the nodes' squared amplitudes and (implicitely) as a function of $\omega$.  
In such a spectrum $\mathcal{A}(\omega)$, resonances appear as pronounced peaks, while regions of consensus are flat. Note that, the value of $\mathcal{A}(0)$ corresponds to the case of no coupling modulation, i.e. the reference model of a static network $\mathbf{A}$. {Without loss of generality} we initialize both models as resting non-consensus states, i.e. we set the state variables at time $t=0$ randomly on the interval $x_i(0)\in\mathcal{I}_\mathrm{init}$ and assume vanishing initial velocities $\dot{x_i}(0)=0, \quad\forall i$. Unless indicated differently we take $\mathcal{I}_\mathrm{init}=[-1,1]$, $h=0.2$ and set  ($d=0.01$, $\gamma=1$) and ($\gamma=1$, $\mu=0.01$) for the DO and SOC model, respectively. {For reasons of comparability between the models, we chose equal values for the damping and velocity alignment coefficients. As a consequence of this choice, the predicted resonance frequencies are similar for both models and it is ensured that the conditions \eqref{eq:cond}-\eqref{eq:cond2} and \eqref{eq:condition1_SOC}-\eqref{eq:condition2_SOC} are satisfied.} In the following we discuss our theoretical results on complete graphs, ring networks and finally on random Erd\H{o}s-R\'enyi networks \cite{erdHos1960evolution}.

\begin{figure}\includegraphics[width=\linewidth]{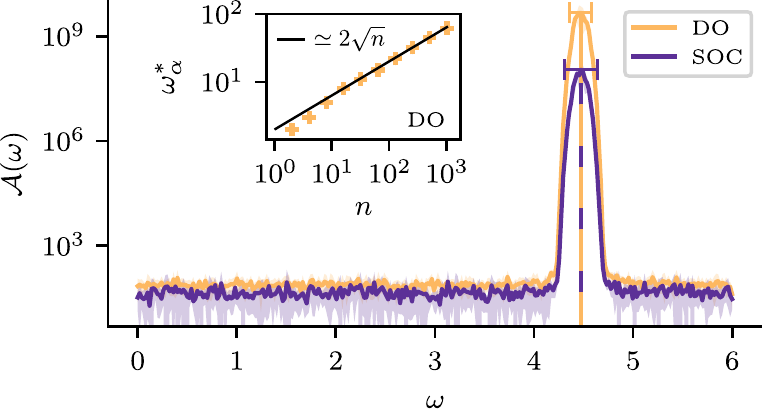}
\caption{\textbf{Resonance on a complete graph.} The vertical orange and dashed magenta lines locate the predicted resonance frequency $\omega^*$ for each model on a complete graph of $n=5$ nodes. The orange and magenta spectra correspond to the DO and SOC model, respectively, on the interval $\omega\in[0,6]$. The spectra are computed as averages over 10 random initial conditions, and the shaded areas show the corresponding standard deviations. The widths of the resonance peaks are approximately captured by the interval of  $\varepsilon$ around $\omega^*$, cf. Eq.~\eqref{eq:epsilon_range} and Eq.~\eqref{eq:s_alpha_SOC}. The inset shows, for the DO model, the theoretically predicted values of $\omega^*$ (black line), and as extracted from simulations (orange crosses) as functions of the system size.} \label{fig:complete_graph} 
\end{figure} 

\subsection*{Amplitude growth}
In a first step we check the theoretical predictions with respect to the exponential growth of the system's amplitudes in the case of resonant modulation. Here, we consider for simplicity the case without both damping ($d=0$) and velocity alignment ($\mu=0$), where the two considered models coincide: the expressions for the resonance frequencies, cf. Eqs.~\eqref{eq:def-res},\eqref{eq:res-ren-attkins}, yield  $\omega_\alpha^*=2\sqrt{\gamma\lambda_\alpha^{(0)}}$, and the exponent $s_\alpha$ becomes 
\begin{equation}\label{eq:s-amp-grow}
    s_\alpha = \frac{h\sqrt{\gamma\lambda_\alpha^{(0)}}}{4}\,.
\end{equation}
The case of such parametric excitation is shown in Fig.~\ref{fig:single_exp_growth} for a complete graph of $n=100$ nodes, where the single resonance is obtained as $\omega^*= 20$. For clarity the trajectory and the local maximum values of a single node $(i=1)$ are shown in green. The trajectories $x_j(t)$ of the remaining agents are colored in grey. 
Clearly, for resonant modulation the nodes' amplitudes increase exponentially over time. The growth is well described by $c_\alpha(t)\sim e^{st}$ (black line), where $s$ is given in Eq.~\eqref{eq:s-amp-grow}.
As the inset of Fig.~\ref{fig:single_exp_growth} demonstrates, the behavior is not only captured for short times, but the approximation also holds over many orders of magnitude for larger times $t\gg1$.

\begin{figure*}
\includegraphics[width=\linewidth]{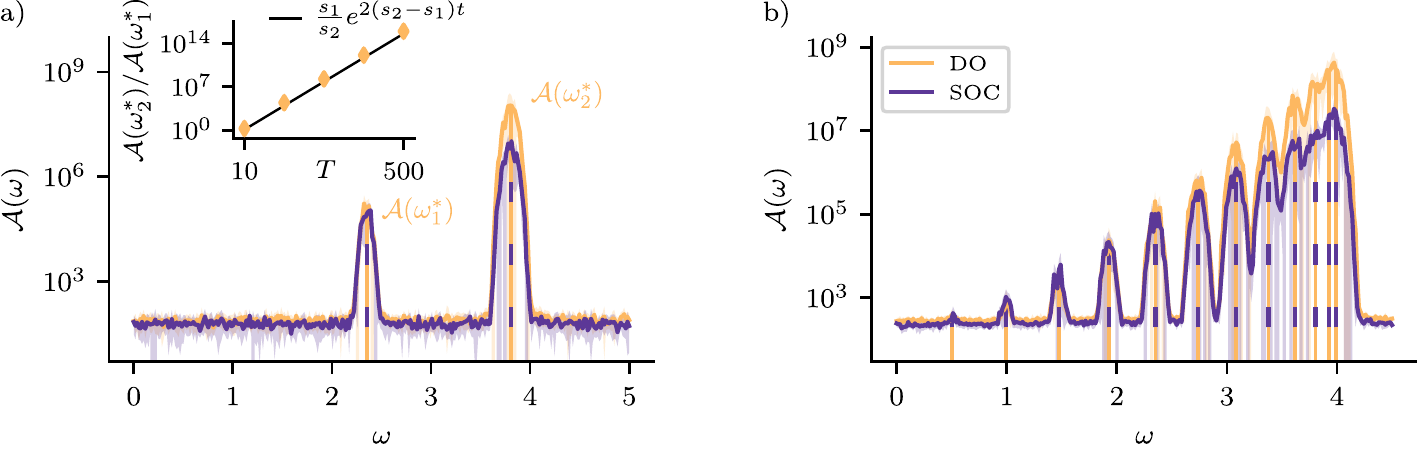}
\caption{\textbf{Resonances on ring networks.} The orange and dashed magenta vertical lines locate the analytically predicted resonance frequencies for both models on a ring network of $n=5$ (a) and $n=25$ (b) nodes, respectively. The resonance spectra are computed as averages over 10 random initial conditions and a simulation time of $T=100$. The inset in panel (a) depicts the ratio $\mathcal{A}(\omega^*_2)/\mathcal{A}(\omega^*_1)$ for the DO model, as a function of the simulation time $T$ (orange rhombuses). The time-dependent ratio is well described by $e^{2(s_2-s_1)t}$ (black line), where $s_1$ and $s_2$ are computed using $\omega_1$ and $\omega_2$, cf. Eq.~\eqref{eq:s}.}\label{fig:ring_network}
\end{figure*}

\subsection*{Complete graphs}
In order to verify the theoretical predictions for the resonance frequencies, we first consider complete graphs, or all-to-all connected networks -- a general topology, often considered in studies of coupled oscillators \cite{Rod16}.
The eigenvalues of $\mathbb{L}^{(0)}$ for a complete graph are given as $\lambda_0=0$ and $\lambda_1=n$, where $\lambda_1$ has the algebraic multiplicity $n-1$ \cite{chung1997spectral}. 

Based on Eqs.~\eqref{eq:def-res},\eqref{eq:omega_SOC} we expect parametric resonance around a single frequency 
\begin{equation}\label{eq:omega_1_complete_graph_A}
    \omega^*=2\sqrt{\gamma n-\left(\frac{d}{2}\right)^2}\,,
\end{equation}
for the DO model and
\begin{equation}\label{eq:omega_1_complete_graph_B}
    \omega^*=2\sqrt{\gamma n-\left(\frac{\mu n}{2}\right)^2}\,.
\end{equation}
in the case of the SOC model, respectively.

In Fig.~\ref{fig:complete_graph} we depict the corresponding resonance spectra on a complete graph of $n=5$ nodes. Here, and in all subsequent figures the orange and magenta lines depict the spectra of the DO and SOC model, respectively. Vertical lines locate the predicted values for $\omega_\alpha^*$, colored according to the associated model. 
For both models, the single peaks of integrated square amplitudes $\mathcal{A}(\omega)$ are predicted remarkably well by the theoretical values of $\omega^*$. Furthermore, the finite widths of the peaks are approximately captured by the theoretical bounds of detuning $\varepsilon$.

The thermodynamic limit of $n\rightarrow\infty$ yields different behaviors for both models. For fixed values of $\mu$ and $\gamma$, no resonance is expected for the SOC model, as the resonance conditions are violated for large $n$. Instead, in the case of the DO model, the damping coefficient does not influence the expression for $\omega^*$ significantly. Hence, the resonance frequency is well described by $\omega^*_\infty\simeq 2\sqrt{n}$, which is supported by numerical simulations for network sizes up to $n=1000$, see inset of Fig.~\ref{fig:complete_graph}(a). For growing system sizes the establishment of a global consensus in the DO model is therefore increasingly robust against low frequency modulations in $f(t)$. Solely high frequency-components may parametrically excite the system and inhibit consensus. 

\subsection*{Rings}
In contrast to the previous case of complete graphs, the Laplacian spectrum of a ring network may contain more than two distinct eigenvalues, which may generally leads to multiple resonances. The set $\{\lambda_\alpha^{(0)}\}$ distributes over the interval $\lambda_\alpha^{(0)}\in[0, 4]$ and the eigenvalues are analytically given as,
\begin{eqnarray}\label{eq:lambda_ring}
\lambda_\alpha^{(0)}=2-2\cos(k_\alpha) \; ,\; \text{with}\; k_\alpha=\frac{2\pi(\alpha-1)}{n} \,.
\end{eqnarray}
This leads to the following expressions for the resonance frequencies:
\begin{equation}\label{eq:omega_ring_A}
\omega_\alpha^*=
2\sqrt{2-2\cos(k_\alpha)-\left(\frac{d}{2}\right)^2}\,
\end{equation}
in the case of the DO model, and like-wise
\begin{equation}\label{eq:omega_ring_B}
\omega_\alpha^*=
2\sqrt{2-\mu^2+(2\mu-2)\cos(k_\alpha)-\mu^2\cos^2(k_\alpha)}\,
\end{equation}
for the SOC model. Note, that in the descriptive case of a ring network of $n=5$ nodes, both distinct eigenvalues $\lambda_\alpha^{(0)}>0$ 
give rise to two valid resonance frequencies $\omega_\alpha^*$, i.e. all resonance conditions are satisfied for $d=\mu=0.01$. 

\begin{figure*}\includegraphics[width=\linewidth]{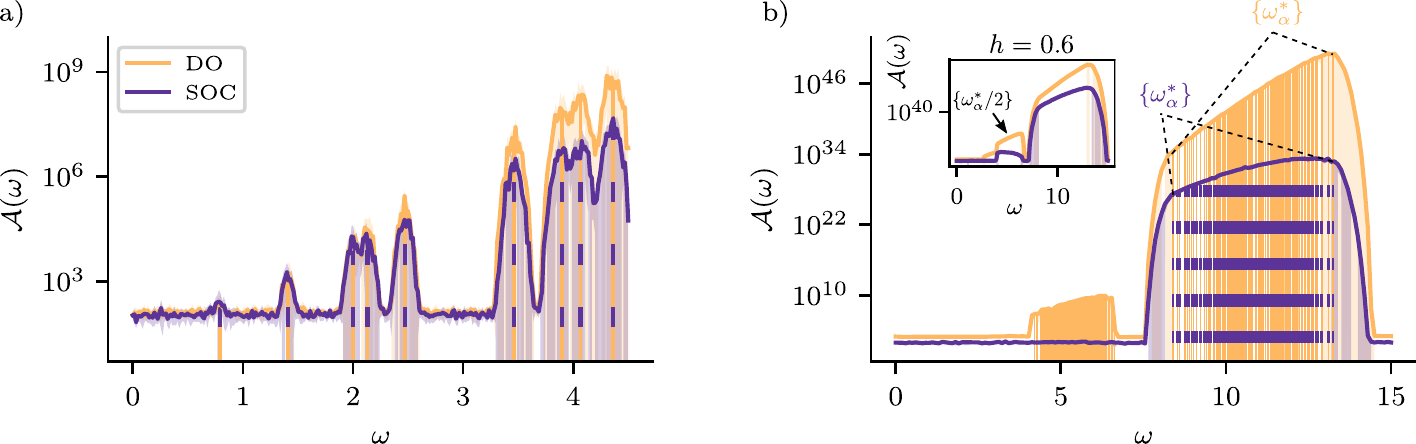} \caption{\textbf{Resonances on Erd\H{o}s-R\'enyi networks.} The orange and dashed magenta vertical lines locate the analytically predicted resonance frequencies $\omega^*_\alpha$ for both models on Erd\H{o}s-R\'enyi networks of $n=25$ (a) and $n=100$ (b) nodes, respectively. The spectra are averaged over 10 random initial conditions, and the shaded blue area shows the corresponding standard deviation. In panel (b) the modulation amplitude is $h=0.4$. The inset in panel (b) shows the spectrum obtained for a larger modulation amplitude $h=0.6$, where half resonant frequencies excite the system also for the SOC model.}
\label{fig:erdos_renyi}
\end{figure*}

The corresponding spectra are depicted in 
Fig.~\ref{fig:ring_network}(a), where both resonances show up as pronounced peaks, matched by their theoretical predictions. For the chosen simulation time of $T=100$, the heights of the resonance peaks, $\mathcal{A}({\omega^*_1})$ and $\mathcal{A}({\omega^*_2})$, are separated by several orders of magnitude, where the separation is larger in the case of the DO model. In the inset of Fig.~\ref{fig:ring_network}(a) we demonstrate, for the DO model, that the ratio of observed peak heights, may also be captured by the theory. It is explained by different exponential growth rates $s_\alpha$, and can, for large $t$, be approximated as $\int c_2 {\rm d}t/\int c_1 {\rm d}t\sim (s_1/s_2)\,e^{2(s_2-s_1)t}$ (black line), obtained from Eq.~\eqref{eq:s}. It captures well the numerical results over many orders of magnitude.

In Fig.~\ref{fig:ring_network}(b) we show the resonance spectra for larger ring networks of $n=25$ nodes, where an important consequence of the bounded Laplacian spectrum becomes apparent. At odds with the behavior of the DO model on complete graphs, increasing the number of nodes for ring networks does not increase the highest resonance frequency beyond all limits. 
Instead, for $n\rightarrow\infty$, the bounded intervals of resonance frequencies 
are gradually filled with resonance peaks, as suggested by Eqs.~\eqref{eq:lambda_ring}. Once two resonant frequencies are sufficiently close to each other, i.e. within their respective $\varepsilon$--ranges, also intermediate frequencies may parametrically excite the system, such that each resonance spectrum becomes a quasi-continuous one. The effect becomes pronounced for higher modulation frequencies $\omega\rightarrow4$, where the increased values of $\mathcal{A}(\omega)$, between two predicted resonance frequencies $\omega_\alpha^*$, indicate a partially off-resonant excitation of the system. 
\subsection*{Random Networks}
Finally we demonstrate that the theory holds on random networks, i.e. that resonance frequencies (based on $\mathbb{L}^{(0)}$) can be predicted reliably on networks with disorder. In the previous cases of regular ring networks and complete graphs, the Laplacian spectrum was given analytically. This is not possible for random networks and the corresponding eigenvalues need to be determined numerically.


In panels (a) and (b) of Fig.~\ref{fig:erdos_renyi} we depict the resonance spectra for two Erd\H{o}s-R\'enyi (ER) networks of $n=10$ and $n=100$ nodes, respectively. Especially, in the former case, the random network structure is revealed by the rather irregular positions of the resonance peaks -- an observation, which is directly related to the random spectrum of the corresponding Laplacian $\mathbb{L}^{(0)}$. Nevertheless, the resonances for both models, are successfully predicted by Eqs.~\eqref{eq:def-res} and \eqref{eq:omega_SOC}. 

In the case of the larger ER network, cf. panel (b) of Fig.~\ref{fig:erdos_renyi}, the resonance frequencies are very close to each other, such that the single vertical lines are barely distinguishable. The overlapping detuning ranges $\epsilon$, hence, lead to a quasi-continuous excitation of the system on the whole interval of roughly $\omega\in[8,13]$ for both models.
Interestingly, the resonance spectrum of the DO model, shows a further feature, known for general parametric oscillators: the excitation of the system at half the resonance frequency \cite{landau1984lehrbuch}. {As we show in the inset of Fig.~\ref{fig:erdos_renyi} for $h=0.6$, this effect becomes increasingly pronounced for larger modulation amplitudes $h$, and is not limited to the DO model, but also appears for the SOC model.}


\section{Summary and Conclusions}\label{sec:conclusions}
In this work we studied parametric resonances emerging in a general model of networked agents, which we discussed in two different settings: i) a model for damped harmonic oscillators, and ii) a well established model for second-order consensus formation, without dissipation and with velocity alignment. We considered those models for time-periodic coupling modulations, and found that, for certain intermediate time scales of such modulation, the collective dynamics is not captured by the time-averaged coupling ${\mathbf{A}}=\langle\mathbf{W}(t)\rangle_t$, which would lead to a global consensus state.
Instead, periodic coupling modulations may induce parametric resonance on a collective level, where single network modes are excited and the emergence of consensus is inhibited. The developed formalism correctly predicts those resonance frequencies, within the theory of parametric resonance, which we extended to the domain of networks with time-varying coupling.
As we demonstrated, resonant modulation gives rise to exponentially growing mode amplitudes, which can be quantified in terms of the associated eigenvalues of the static Laplacian $\mathbb{L}^{(0)}$. In this context, both dissipation and velocity alignment counteract the excitation of the system and therefore foster the establishment of global consensus states. 

In comparison both considered models, DO and SOC, are similarly susceptible to time-periodic coupling modulations. The excitation for a fixed resonance frequency, however, is increased for the DO model, as suggested by the numerically obtained resonance spectra. For small system sizes and rather weak dissipation, as well as velocity alignment, the expected resonance frequencies are in close vicinity. Using the example of complete graphs, we illustrated that this generally changes for large networks. Increasingly large eigenvalues would not give rise to resonances in the SOC model, which emphasized the enhanced positive effect on consensus formation for velocity alignment, compared to dissipation, cf. condition~\eqref{eq:condition2_SOC}. 

Interestingly, we found that the formalism also holds on networks of non-linearly coupled second-order Kuramoto oscillators close to a synchronization fixed point. It correctly predicts modulation frequencies, for which the synchronized fixed point becomes unstable. This did not yield exponentially growing but finite oscillation amplitudes. 

Our work might have important implications for the robustness of second-order consensus algorithms applied in real systems. For example, in order to disrupt consensus states within networked vehicular platoons or cooperating autonomous robots, signals which periodically weaken the units' sensors, may be used. Note that parametric resonance might also be used to infer characteristic networks features, in cases where the probing of the system is not localized on certain nodes, but performed on the overall coupling.




\section*{Acknowledgments}
This work was developed within the
scope of the IRTG 1740/TRP 2015/50122-0 and funded
by the DFG/FAPESP and the Swiss National Science Foundation under grant No. 200020\_\!\_182050.
We thank Adrian Pacheco, J\"org N\"otel and Thomas Peron for fruitful discussions and helpful comments on the manuscript.

\appendix
\section{Non-linear oscillators}\label{sec:non-lin-osc}
In the main text we considered (linear) diffusively coupled second-order consensus systems. 
However, Eq.~\eqref{eq:vectorial} may also be regarded as the linearization of non-linearly coupled oscillators.

Traditionally, one of the most studied models of non-linearly coupled oscillators is the Kuramoto model \cite{kuramoto1975self}. Second-order Kuramoto models were previously used to investigate synchronization phenomena, such as particular cases of consensus formation~\cite{Pat10} or frequency synchronization in power grids~\cite{Ber81}. 
By exchanging the diffusive coupling by a sinusoidal coupling, setting $\mu=0$ and switching to the common notation for phase oscillators, i.e. $x_i\rightarrow\phi_i$, Eq.~(\ref{eq:vectorial}) yields the second-order Kuramoto model as
\begin{eqnarray}\label{eq:kura}
\ddot{\phi}_i + d\,\dot{\phi}_i = \omega_i - \sum_{j=1}^{N} W_{ij}(t)\sin(\phi_i-\phi_j)\;.
\end{eqnarray}
At odds with the linear model of Eq.~(\ref{eq:vectorial}) the boundedness of the sinusoidal coupling inhibits an arbitrarily large excitation of the system in the case of resonant modulation. Instead, 
the nodes' trajectories $x_i$ are characterized by large but bounded oscillation amplitudes. 

Without loss of generality, we take $\sum_i\omega_i=0$ which yields a synchronous state with $\dot{\phi}_i=0$ $\forall i$\,. Furthermore we assume that Eq.~(\ref{eq:kura}) has a stable fixed point $\phi_{i,0}$ for a specific backbone network ${\mathbf{A}}$, which yields
\begin{eqnarray}
\omega_i=\sum_{j=1}^{N} W_{ij}(t)\sin(\phi_{i,0}-\phi_{j,0})\;. 
\end{eqnarray}
For small modulation amplitudes in $f(t)$, i.e. $h\ll1$ the dynamics can be linearized around the stable fixed point. Inserting $\delta\phi_{i}(t)=\phi_i(t)-\phi_{i,0}$ into Eq.~(\ref{eq:kura}) yields
\begin{align}
\begin{split}
\delta\ddot{\phi}_i + d\,\delta\dot{\phi}_i = -\sum_{j=1}^{N} W_{ij}(t)\cos(\phi_{i,0}-\phi_{j,0})(\delta\phi_{i,0}-\delta\phi_{j,0})\;.
\end{split}
\end{align}
In vectorial form we have
\begin{eqnarray}
\delta\Ddot{\bm \phi} +  d\delta\dot{\bm \phi} = -\mathbb{L}(t;{\bm \phi}_0)\,\delta{\bm\phi}\,,
\end{eqnarray}
where $\mathbb{L}(t;{\bm \phi}_0)$ denotes a Laplacian matrix, weighted by the angle differences at the fixed point ${\bm \phi}_0$. Therefore, the relations derived in Sec.~\ref{sec1} still hold -- where $\{\lambda_\alpha^{(0)}\}$ now denotes the eigenvalue spectrum of the weighted Laplacian $\mathbb{L}(t;{\bm \phi}_0)$.

Note, that we focus on deviations from a global consensus for which holds $\phi_{0,i}=\phi_{0,j},\,\forall i\neq j$. In this case the weighted Laplacian reduces to the one used in the linear model [Eq.~\eqref{eq:vectorial}], i.e. we have $\mathbb{L}(t, {\bm\phi}_0)=\mathbb{L}(t)$.

\begin{figure}\includegraphics[width=0.98\linewidth]{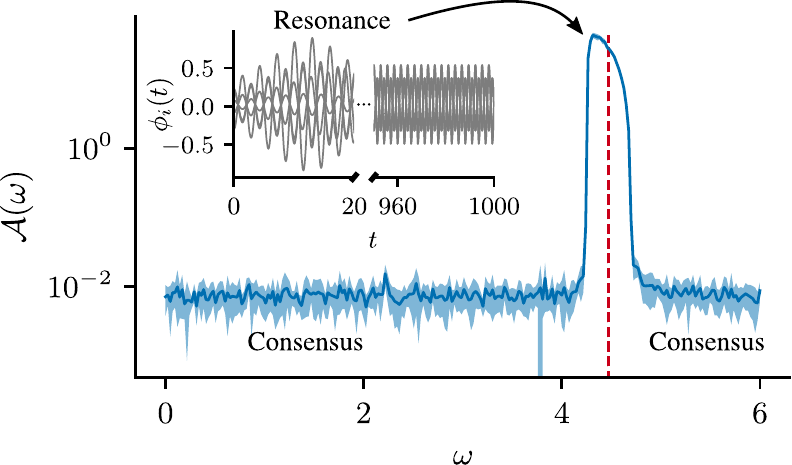}  \caption{\textbf{Resonance for the second-order Kuramoto model.} The red line locates the predicted resonance frequency $\omega^*$ for in the case of a complete graph of $n=5$ nodes close to the consensus fixed point. The blue line depicts the resonance spectrum for the second-order Kuramoto model on the interval $\omega\in[0,6]$, averaged over 10 random initial conditions, and the shaded blue area shows the standard deviation. The initial interval was set to $\mathcal{I}_\mathrm{init}=[-0.01,0.01]$. The inset shows the time evolution of $\phi_i(t)$ for resonant modulation with $\omega^*\simeq 4.3$.}
\label{fig:kuramoto}
\end{figure}

The inset of Fig.~\ref{fig:kuramoto} shows the time evolution of the agents' phases $\phi_i$ for resonant modulation ($\omega=\omega^*$) on a complete graph of $n=5$ nodes.  
As expected, in the case of resonance, consensus is inhibited. In contrast to the linear models (DO and SOC), however, oscillation amplitudes do not grow exponentially without bounds. Instead, phase oscillations reach finite amplitudes, which would vanish for off-resonant modulation,
where agents will reach a global consensus, as in the linear case, depicted in Fig.~\ref{fig:fig1_example}.
This behavior is consistent for the whole range of relevant modulation frequencies as shown in Fig.~\ref{fig:kuramoto}, where a pronounced peak appears for the same frequency as predicted for the DO model, cf. Fig.~\ref{fig:complete_graph}. Note, that the linearization of the sinusoidal coupling only holds for the dynamics close to the fixed point of consensus. To observe resonance in the numerical simulations, the Kuramoto model needs therefore to be probed close to this fixed point ($|\phi_i(0)-\phi_j(0)|\ll1$ and $|\dot{\phi}
_i(0)-\dot{\phi}_j(0)|\ll1 \,\forall i\neq j$), for which we decreased the width of the initial interval of $\phi_i(0)$ to  $\mathcal{I}_\mathrm{init}=[-0.01,0.01]$.

\end{document}